# Pareto Curves for Probabilistic Model Checking

Vojtěch Forejt[1], Marta Kwiatkowska[1], and David Parker[2]

[1] Department of Computer Science, University of Oxford, UK
[2] School of Computer Science, University of Birmingham, UK

**Abstract.** Multi-objective probabilistic model checking provides a way to verify several, possibly conflicting, quantitative properties of a stochastic system. It has useful applications in controller synthesis and compositional probabilistic verification. However, existing methods are based on linear programming, which limits the scale of systems that can be analysed and makes verification of time-bounded properties very difficult. We present a novel approach that addresses both of these shortcomings, based on the generation of successive approximations of the Pareto curve for a multi-objective model checking problem. We illustrate dramatic improvements in efficiency on a large set of benchmarks and show how the ability to visualise Pareto curves significantly enhances the quality of results obtained from current probabilistic verification tools.

## 1 Introduction

Probabilistic model checking is an automated technique for verifying systems that exhibit stochastic behaviour. This arises due to, for example, failures in physical components, unreliable communication media or randomisation. Systems are typically modelled as Markov chains or Markov decision processes (MDPs), and probabilistic temporal logics are used to specify quantitative properties to be verified such as "the probability of a message packet being lost is less than 0.05" or "the expected energy consumption is at most 100 mJ".

It is often necessary to incorporate *nondeterminism* into system models, to represent, for example, the actions of an external controller or the order in which a scheduler chooses to interleave system components running in parallel. In these cases, systems are usually modelled as MDPs. Each possible way of resolving the nondeterminism in an MDP is represented by an *adversary* (also known as a strategy or policy). Properties to be verified against the MDP quantify over its adversaries, e.g., "the probability of a message packet being lost is less than 0.05 *for all possible adversaries*". It is also common to use numerical queries, e.g., "what is the *maximum* expected energy consumption?".

Model checking reduces to an *optimisation* problem, namely determining the maximum (or minimum) probability (or expected cost/reward) achievable by any adversary. For the most common classes of property (the probability of reaching a set of target states or the expected cumulated reward), model checking can be reduced to solving a linear programming (LP) problem. In practice, however, most probabilistic verification tools use (approximate) iterative numerical methods, such as *value iteration* [19], since they scale to much larger systems and are amenable to symbolic (BDD-based) implementations. Value iteration can also be used for *time-bounded* (finite-horizon) properties, which is impractical with LP. Another alternative is policy iteration, but this is also impractical for time-bounded properties, and preliminary investigations in [10] showed no particular improvement over value iteration in the context of probabilistic verification.

There has recently been increased interest in *multi-objective* probabilistic model checking for MDPs [5,9,11,4], which can be used to analyse trade-offs between several, possibly conflicting, quantitative properties. Consider, for example, two events of interest, $A$ and $B$, and let $p_A^\sigma$ and $p_B^\sigma$ be the probability that each occurs under an adversary $\sigma$ of an MDP. In this paper, we study several kinds of multi-objective properties. *Achievability* queries ask, e.g., "is there an adversary $\sigma$ satisfying the predicate $\boldsymbol{\psi} = p_A^\sigma \geqslant x \wedge p_B^\sigma \geqslant y$?" and *numerical* queries ask, e.g., "what is maximum value of $x$ such that $\boldsymbol{\psi}$ is achievable?". We also consider the *Pareto curve* of *undominated* solution points: for this example, the set of pairs $(x, y)$ such that $\boldsymbol{\psi}$ is achievable but any increase in either $x$ or $y$ would necessitate a decrease in the other.

Multi-objective techniques have natural applications to *controller synthesis* for MDPs (e.g., "how can we maximise the probability of successful message transmission, whilst keeping the expected energy usage below 100 mJ?"). They also form the basis of recent *compositional verification* techniques [15], which decompose model checking into separate tasks for each system component using assume-guarantee reasoning (e.g., "what is the maximum probability of a global system error, under the assumption that component 1 fails with probability at most 0.02?"). This approach has been successfully used to verify probabilistic systems too large to analyse without compositional techniques.

Existing multi-objective model checking methods [5,9,11,4] rely on a reduction to LP. The linear program solved, although of a rather different form to the standard (single objective) case, is still linear in the size of the MDP, yielding polynomial time complexity. As discussed above, though, LP-based probabilistic verification has several important weaknesses. In this paper, we present a novel approach to multi-objective model checking of probabilistic reachability and expected total reward properties. Our method is based on the generation of successive, increasingly precise approximations to the Pareto curve by optimising weighted sums of objectives using value iteration. On a large selection of benchmarks, we demonstrate the following benefits:

(i) dramatic improvements in run-time *efficiency*, by factors of up to 150;
(ii) significant *scalability* improvements: over an order of magnitude model size;
(iii) the usefulness of visualising *Pareto curves* for verification problems;
(iv) solution of *time-bounded* probabilistic reachability and cumulative reward.

The last of these also paves the way for the development of multi-objective techniques for richer, *timed* classes of models such as continuous-time MDPs.

This paper is an extended version, with additional details and proofs, of [12].

**Related work.** Multi-objective optimisation has been extensively studied in areas such as operations research, economics and stochastic control [6], including its application to MDPs [1]. Many general approaches exist, based on, for example, *normalising* multiple objectives into a single weighted objective; *constrained* approaches optimising one objective while bounding the others; heuristic search using, e.g., *evolutionary algorithms* [7]; application of satisfiability/constraint solvers [17]; and stochastic search with restarts [16]. Several methods, including e.g. [16], iterate over weighted sums of objectives, as we do; the main difference is that our approach is tailored to the convex, linear problems derived from MDPs.

Multi-objective optimisation is routinely used in areas such as embedded systems design and a variety of general-purpose optimisation tools exist, e.g., PISA [2], ModeFRONTIER and libraries for MATLAB. Such tools tend to be targeted at much more complex (e.g., non-convex and non-linear) design spaces than the ones that we focus on in this paper. They are also typically used for *static* design problems, rather than our *dynamic* models of system behaviour.

A rigorous complexity analysis of multi-objective optimisation, in particular for approximating *Pareto curves*, was undertaken in the influential work of [18]. More recently [8], some of these results were improved for the simpler case of *convex* multi-objective problems but no practical investigations are undertaken.

Most relevant to the current work is the application of multi-objective optimisation to *probabilistic verification* [5,9,11,4]. In [5,9,4], discounted total reward, probabilistic $\omega$-regular and long-run average properties are studied, respectively. In each case, algorithms are given using a reduction to LP, also showing the existence of methods to approximate Pareto curves using the results of [18], but implementations are not considered. The work of [11] adds expected total reward properties and provides an implementation, based on LP. As discussed above, the performance and scalability of our approach is significantly better, as we show in Section 5. None of the above consider time-bounded properties. In principle, for discrete-time models like MDPs, these can be reduced to unbounded properties using a finite counter but this is generally impractical in terms of scalability.

## 2 Background

**Geometry.** For a vector $\boldsymbol{x} \in \mathbb{R}^n$, we use $x_i$ to denote its $i$-th component and say $\boldsymbol{x}$ is a *weight vector* if $x_i \geqslant 0$ for all $i$ and $\sum_{i=1}^n x_i = 1$. The *Euclidean inner product* of $\boldsymbol{x}, \boldsymbol{y} \in \mathbb{R}^n$ is defined as $\boldsymbol{x} \cdot \boldsymbol{y} = \sum_{i=1}^n x_i \cdot y_i$. For a set of vectors $X = \{\boldsymbol{x}_1, \ldots \boldsymbol{x}_k\} \subseteq \mathbb{R}^n$, a *convex combination* is $\sum_{j=1}^k w_j \cdot \boldsymbol{x}_j$ for some weight vector $\boldsymbol{w} \in \mathbb{R}^k$. We use $down(X)$ to denote the *downward closure* of the convex hull of $X$, i.e. the set of vectors $\boldsymbol{z} \in \mathbb{R}^n$ that satisfy $\boldsymbol{z} \leqslant \boldsymbol{y}$ for some convex combination $\boldsymbol{y}$ of $X$. Given a convex set $Y$, we say that a point $\boldsymbol{y} \in Y$ is on the *boundary* of $Y$ if, for any $\varepsilon > 0$, there is a point $\boldsymbol{z} \notin Y$ such that the Euclidean distance between $\boldsymbol{y}$ and $\boldsymbol{z}$ is at most $\varepsilon$. From the *separating hyperplane* and *supporting hyperplane* theorems, we have the following.

**Proposition 1 ([3]).** *Let $Y \subseteq \mathbb{R}^n$ be a downward closed set of points. For any $\boldsymbol{p} \in \mathbb{R}^n$ not in $Y$, there is a weight vector $\boldsymbol{w} \in \mathbb{R}^n$ such that $\boldsymbol{w} \cdot \boldsymbol{p} > \boldsymbol{w} \cdot \boldsymbol{y}$ for all $\boldsymbol{y} \in Y$. Also, for any $\boldsymbol{q}$ on the boundary of $Y$, there is a weight vector $\boldsymbol{w} \in \mathbb{R}^n$ such that $\boldsymbol{w} \cdot \boldsymbol{q} \geqslant \boldsymbol{w} \cdot \boldsymbol{y}$ for all $\boldsymbol{y} \in Y$. We say that $\boldsymbol{w}$* separates $\boldsymbol{q}$ from $down(Y)$.

**Markov decision processes (MDPs).** MDPs are commonly used to model systems with probabilistic and nondeterministic behaviour. Denoting by $Dist(X)$ the set of probability distributions over a set $X$, an *MDP* takes the form $\mathcal{M} = (S, \bar{s}, \alpha, \delta)$, where $S$ is a set of states, $\bar{s} \in S$ is an initial state, $\alpha$ is a set of actions and $\delta : S \times \alpha \to Dist(S)$ is a (partial) probabilistic transition function.

Each state $s$ of an MDP $\mathcal{M}$ has an associated set $A(s)$ of *enabled* actions, given by $A(s) \stackrel{\text{def}}{=} \{a \in \alpha \mid \delta(s, a) \text{ is defined}\}$. If action $a \in A(s)$ is taken in state $s$, then the next state is determined randomly according to the distribution $\delta(s, a)$, i.e., a transition to state $s'$ occurs with probability $\delta(s, a)(s')$. A *path* through $\mathcal{M}$ is a (finite or infinite) sequence $\pi = s_0 a_0 s_1 a_1 \dots$ where $s_0 = \bar{s}$, and $a_i \in A(s_i)$ and $\delta(s_i, a_i)(s_{i+1}) > 0$ for all $i$. We denote by *IPaths* (*FPaths*) the set of all infinite (finite) paths and, for finite path $\pi$, $last(\pi)$ is its last state.

An *adversary* $\sigma : FPaths \to Dist(\alpha)$ (also called a strategy or policy) of $\mathcal{M}$ is a resolution of the choices of action in each state, based on its execution so far. In standard fashion [13], an adversary $\sigma$ induces a probability measure $Pr^{\sigma}_{\mathcal{M}}$ over *IPaths*. An adversary $\sigma$ is *deterministic* if $\sigma(\pi)$ is a Dirac distribution for all $\pi$ (and *randomised* if not); it is *memoryless* if $\sigma(\pi)$ depends only on $last(\pi)$. The set of all adversaries for $\mathcal{M}$ is $Adv_{\mathcal{M}}$.

A *reward structure* for $\mathcal{M}$ is a function $\rho : S \times \alpha \to \mathbb{R}$ mapping actions to (positive or negative) reals. For an infinite path $\pi = s_0 a_0 s_1 a_1 \dots$ and a number $k \in \mathbb{N} \cup \{\infty\}$ the *total reward in $k$ steps* for $\pi$ over $\rho$ is $\rho[k](\pi) \stackrel{\text{def}}{=} \sum_{i=0}^{k-1} \rho(s_i, a_i)$.

**Model checking MDPs.** In this paper, we focus on two key classes of properties for MDPs: the *probability of reaching a target* and the *expected total reward*. In each case, we consider both time-bounded and unbounded variants. We will later discuss generalisation to more expressive properties. In this and the following sections, we assume a fixed MDP $\mathcal{M} = (S, \bar{s}, \alpha, \delta)$.

**Definition 1 (Reachability predicate).** *A* reachability predicate $[T]^{\leqslant k}_{\sim p}$ *comprises a set of target states $T \subseteq S$, a relational operator $\sim \in \{\geqslant, \leqslant\}$, a rational probability bound $p$ and a time-bound $k \in \mathbb{N} \cup \{\infty\}$. It states that the probability of reaching $T$ within $k$ steps satisfies $\sim p$. Formally, satisfaction of $[T]^{\leqslant k}_{\sim p}$ by MDP $\mathcal{M}$, under adversary $\sigma$, denoted $\mathcal{M}, \sigma \models [T]^{\leqslant k}_{\sim p}$, is defined as follows:*

$$\mathcal{M}, \sigma \models [T]^{\leqslant k}_{\sim p} \Leftrightarrow Pr^{\sigma}_{\mathcal{M}}(\{s_0 a_0 s_1 a_1 \dots \in IPaths \mid \exists i \leqslant k \, : \, s_i \in T\}) \sim p \, .$$

**Definition 2 (Reward predicate).** *A* reward predicate $[\rho]^{\leqslant k}_{\sim r}$ *comprises a reward structure $\rho : S \times \alpha \to \mathbb{R}$, a relational operator $\sim \in \{\geqslant, \leqslant\}$, a rational reward bound $r$ and a time bound $k \in \mathbb{N} \cup \{\infty\}$. It states that the* expected total reward *cumulated within $k$ steps satisfies $\sim r$. Formally, satisfaction of $[\rho]^{\leqslant k}_{\sim r}$ by $\mathcal{M}$, under adversary $\sigma$, denoted $\mathcal{M}, \sigma \models [\rho]^{\leqslant k}_{\sim r}$, is defined as follows:*

$$\mathcal{M}, \sigma \models [\rho]^{\leqslant k}_{\sim r} \;\Leftrightarrow\; ExpTot^{\sigma,k}_{\mathcal{M}}(\rho) \sim r \;\; \text{where} \;\; ExpTot^{\sigma,k}_{\mathcal{M}}(\rho) \;\stackrel{\text{def}}{=}\; \int_{\pi} \rho[k](\pi) \, dPr^{\sigma}_{\mathcal{M}} .$$

For the *unbounded* forms of the notation above ($k = \infty$), we will often omit $k$, writing e.g. $[\rho]_{\sim r}$ instead of $[\rho]^{\leqslant \infty}_{\sim r}$ or $ExpTot^{\sigma}_{\mathcal{M}}(\rho)$ instead of $ExpTot^{\sigma,\infty}_{\mathcal{M}}(\rho)$.

For this paper, we also need to consider *weighted sums* of rewards.

**Definition 3 (Weighted reward sum).** *Given a weight vector $\boldsymbol{w} \in \mathbb{R}^n$ and vectors of time bounds $\boldsymbol{k} = (k_1, \ldots, k_n) \in (\mathbb{N} \cup \{\infty\})^n$ and reward structures $\boldsymbol{\rho} = (\rho_1, \ldots, \rho_n)$ for MDP $\mathcal{M}$, the* weighted reward sum $\boldsymbol{w}\cdot\boldsymbol{\rho}[\boldsymbol{k}]$ *over a path $\pi$ is defined as $\boldsymbol{w}\cdot\boldsymbol{\rho}[\boldsymbol{k}](\pi) \stackrel{\text{def}}{=} \sum_{i=1}^{n} w_i \rho_i[k](\pi)$. The* expected total weighted sum *is then: $ExpTot^{\sigma,\boldsymbol{k}}_{\mathcal{M}}(\boldsymbol{w}\cdot\boldsymbol{\rho}) \stackrel{\text{def}}{=} \int_\pi \boldsymbol{w}\cdot\boldsymbol{\rho}[\boldsymbol{k}](\pi)\, dPr^\sigma_{\mathcal{M}}$. For any adversary $\sigma$, we have: $ExpTot^{\sigma,\boldsymbol{k}}_{\mathcal{M}}(\boldsymbol{w}\cdot\boldsymbol{\rho}) = \sum_{i=1}^{n} w_i ExpTot^{\sigma,k_i}_{\mathcal{M}}(\rho_i)$.*

Notice that satisfaction of reachability and reward predicates is defined above with respect to a specific adversary $\sigma$ of an MDP $\mathcal{M}$. When performing model checking on the MDP, the most common approach is to verify that such a predicate is satisfied *for all* adversaries $\sigma \in Adv_{\mathcal{M}}$. An alternative, often described as *controller synthesis*, is to ask the dual question: whether *there exists* an adversary $\sigma$ satisfying the predicate. In either case, model checking reduces to computing the maximum or minimum reachability probability or expected reward. For the unbounded cases, this can be done by solving an LP problem, using policy iteration, or with value iteration, an approximate iterative numerical method [19]. For time-bounded properties, only value iteration is applicable.

## 3 Multi-objective Queries

We now describe how to formalise multi-objective queries for MDPs. In the following section, we will present novel, efficient algorithms for their verification. We formulate our queries in a similar style to the one taken in [11], but with two key additions. Firstly, we include the ability to specify *time-bounded* reachability and reward properties. Secondly, we consider *Pareto curves*.

The essence of multi-objective properties for MDPs is that they require multiple predicates to be satisfied concurrently for the same adversary.

**Definition 4 (Multi-objective predicate).** *A* multi-objective predicate *is a vector $\boldsymbol{\psi}=(\psi_1, \ldots, \psi_n)$ of reachability or reward predicates. We say that $\boldsymbol{\psi}$ is* satisfied *by MDP $\mathcal{M}$ under adversary $\sigma$, denoted $\mathcal{M}, \sigma \models \boldsymbol{\psi}$, if $\mathcal{M}, \sigma \models \psi_i$* for all $1 \leqslant i \leqslant n$. *We call $\boldsymbol{\psi}$ a* basic multi-objective predicate *if it is of the form $([\rho_1]^{\leqslant k_1}_{\geqslant r_1}, \ldots, [\rho_n]^{\leqslant k_n}_{\geqslant r_n})$, i.e. it comprises only lower-bounded reward predicates.*

We define three ways to formulate multi-objective queries for an MDP: *achievability queries*, which check for the existence of an adversary satisfying a multi-objective predicate $\boldsymbol{\psi}$; *numerical queries*, which maximise or minimise a reachability/reward objective over the set of adversaries satisfying $\boldsymbol{\psi}$; and *Pareto queries*, which determine the Pareto curve for a set of objectives.

**Definition 5 (Achievability query).** *For MDP $\mathcal{M}$ and multi-objective predicate $\boldsymbol{\psi}$, an* achievability query *asks if $\boldsymbol{\psi}$ is* satisfiable *(or* achievable*), i.e. whether there exists an adversary $\sigma \in Adv_{\mathcal{M}}$ such that $\mathcal{M}, \sigma \models \boldsymbol{\psi}$.*

**Definition 6 (Numerical query).** *For MDP $\mathcal{M}$, a* numerical query *is of the form $num([o_1]^{\leqslant k_1}_{\star}, (\psi_2 \ldots, \psi_n))$, comprising an $n{-}1$-sized multi-objective predicate $(\psi_2 \ldots, \psi_n)$ and an* objective *$[o_1]^{\leqslant k_1}_{\star}$, where $o_1$ is a reward structure $\rho_1$ or*

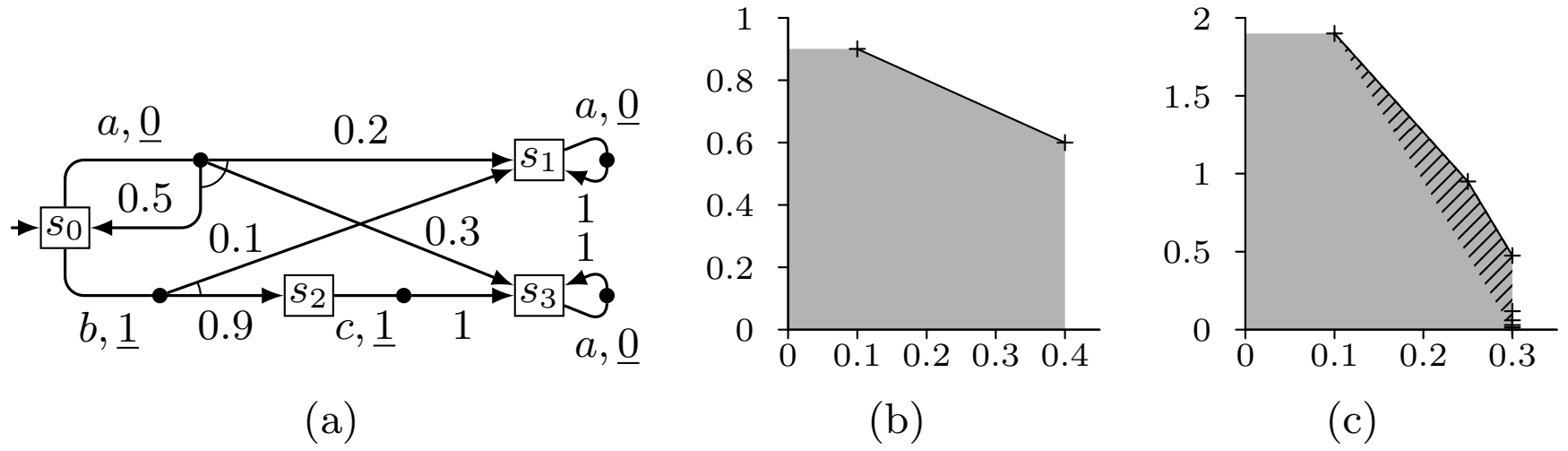

**Fig. 1.** Example MDP (a), graphs for $([\{s_1\}]_{\geqslant x}, [\{s_3\}]_{\geqslant y})$ (b) and $([\{s_1\}]^{\leqslant 2}_{\geqslant x}, [\rho]_{\geqslant y})$ (c).

*target set* $T_1$*,* $k_1 \in \mathbb{N} \cup \{\infty\}$ *is a time bound and* $\star \in \{\min, \max\}$*. We define:*

$$num([o_1]^{\leqslant k_1}_{\min}, (\psi_2, \dots, \psi_n)) \stackrel{\text{def}}{=} \inf\{x \in \mathbb{R} \mid ([o_1]^{\leqslant k_1}_{\leqslant x}, \psi_2, \dots, \psi_n) \textit{ is satisfiable}\}.$$
$$num([o_1]^{\leqslant k_1}_{\max}, (\psi_2, \dots, \psi_n)) \stackrel{\text{def}}{=} \sup\{x \in \mathbb{R} \mid ([o_1]^{\leqslant k_1}_{\geqslant x}, \psi_2, \dots, \psi_n) \textit{ is satisfiable}\}.$$

**Definition 7 (Pareto query).** *For MDP* $\mathcal{M}$*, a* Pareto query *takes the form* $pareto([o_1]^{\leqslant k_1}_{\star_1}, \dots, [o_n]^{\leqslant k_n}_{\star_n})$*, where each* $[o_i]^{\leqslant k_i}_{\star_i}$ *is an objective as in Defn. 6. The set of* achievable values *is* $A = \{\boldsymbol{x} \in \mathbb{R}^n \mid ([o_1]^{\leqslant k_1}_{\sim_1 x_1}, \dots, [o_n]^{\leqslant k_n}_{\sim_n x_n}) \textit{ is satisfiable}\}$ *where* $\sim_i = \geqslant$ *if* $\star_i = \max$ *and* $\sim_i = \leqslant$ *if* $\star_i = \min$*. We say, for points* $\boldsymbol{x}, \boldsymbol{y} \in A$*, that* $\boldsymbol{x}$ dominates $\boldsymbol{y}$ *if* $x_i \sim_i y_i$ *for all* $i$ *and* $x_j \neq y_j$ *for some* $j$*. Then:*

$$pareto([o_1]^{\leqslant k_1}_{\star_1}, \dots, [o_n]^{\leqslant k_n}_{\star_n}) \stackrel{\text{def}}{=} \{\boldsymbol{x} \in A \mid \boldsymbol{x} \textit{ is not dominated by any } \boldsymbol{y} \in A\}.$$

**Convexity.** A fundamental property of the multi-objective optimisation problems solved in this paper (and on MDPs in general) is their *convexity*. More precisely, consider target sets $T_1, \dots, T_n$, reward structures $\rho_1, \dots, \rho_m$ and time-bounds $k_1, \dots, k_n, l_1, \dots, l_m \in \mathbb{N} \cup \{\infty\}$. Let $\boldsymbol{x}^\sigma \in \mathbb{R}^{n+m}$ be the vector defined such that $x_i = Pr^\sigma_{\mathcal{M}}(\Diamond^{\leqslant k_i} T_i)$ for $1 \leqslant i \leqslant n$ and $x_{n+j} = ExpTot^{\sigma,k_j}_{\mathcal{M}}(\rho_j)$ for $1 \leqslant j \leqslant m$, where $Pr^\sigma_{\mathcal{M}}(\Diamond^{\leqslant k_i} T_i)$ denotes the probability of reaching $T_i$ in $k_i$ steps under $\sigma$. Then, the set $\{\boldsymbol{x}^\sigma \mid \sigma \in Adv_{\mathcal{M}}\}$ forms a convex polytope [9,11][1]. As a direct consequence of this, the set of *achievable values* for a Pareto query is also convex.

**Example 1.** Fig. 1(a) shows an MDP with accompanying reward structure $\rho$ indicated by underlined numbers. Consider first the multi-objective predicate $\boldsymbol{\psi} = ([\{s_1\}]_{\geqslant x}, [\{s_3\}]_{\geqslant y})$, which imposes lower bounds on the probabilities of reaching states $s_1$ and $s_3$. The grey area in Fig. 1(b) shows the values of $x$ and $y$ for which $\boldsymbol{\psi}$ is satisfiable. The two points on the graph marked as + correspond to the two possible *memoryless deterministic* adversaries in $\mathcal{M}$. The line joining them (their convex closure) represents the points for all possible adversaries. For this example, this line also constitutes the Pareto curve. Achievability queries on $\boldsymbol{\psi}$ for $(x, y) = (0.2, 0.7)$ and $(0.4, 0.7)$ return true and false, respectively. Numerical query $num([\{s_1\}]_{\max}, ([\{s_3\}]_{\geqslant 0.7}))$ returns 0.3.

Consider a second predicate $\boldsymbol{\psi}' = ([\{s_1\}]^{\leqslant 2}_{\geqslant x}, [\rho]_{\geqslant y})$, now with a time-bounded reachability and reward predicate. Fig. 1(c) depicts (by +) points for some of the

[1] Strictly speaking, this requires finiteness of rewards, which we discuss below.

*deterministic* adversaries of $\mathcal{M}$, of which there are infinitely many. Their convex combination, the dashed area, marks the points achievable by *all* (randomised) adversaries, and its downward closure, in grey, shows the values of $x$ and $y$ for which $\boldsymbol{\psi}'$ is satisfiable. The Pareto curve is the black line along the top edge.

**Assumptions.** For the purposes of model checking the queries described in this section, we need to impose certain restrictions on the use of rewards. For clarity, we describe these in terms of achievability queries but they apply to all three classes. We first need the following definition.

**Definition 8 (Reward-finiteness).** *Let $\mathcal{M}$ be an MDP and consider an achievability query $\boldsymbol{\psi}{=}([T_1]^{\leqslant k_1}_{\sim_1 p_1}, \ldots, [T_n]^{\leqslant k_n}_{\sim_n p_n}, [\rho_1]^{\leqslant l_1}_{\bowtie_1 r_1}, \ldots, [\rho_m]^{\leqslant l_m}_{\bowtie_m r_m})$ for $\mathcal{M}$. We say that $\boldsymbol{\psi}$ is* reward-finite *if, for each $1{\leqslant}j{\leqslant}m$ such that $l_j{=}\infty$ and $\bowtie_j \; = \; \geqslant$, we have:* $\sup\{ExpTot^{\sigma,l_j}_{\mathcal{M}}(\rho_j) \mid \mathcal{M},\sigma \models ([T_1]^{\leqslant k_1}_{\sim_1 p_1}, \ldots, [T_n]^{\leqslant k_n}_{\sim_n p_n})\} < \infty$ *and is* fully reward-finite *if, for each $1{\leqslant}j{\leqslant}m$ such that $l_j{=}\infty$ and $\bowtie_j \; = \; \geqslant$, we have:* $\sup\{ExpTot^{\sigma,l_j}_{\mathcal{M}}(\rho_j) \mid \sigma \in Adv_{\mathcal{M}}\} < \infty$.

Let $\mathcal{M}$ and $\boldsymbol{\psi}$ be as in Defn. 8. To model check $\boldsymbol{\psi}$ on $\mathcal{M}$, we require that: (i) each reward structure $\rho_i$ assigns only non-negative values; (ii) $\boldsymbol{\psi}$ is reward-finite; and (iii) for indices $1{\leqslant}j{\leqslant}m$ such that $l_j{=}\infty$, either all $\bowtie_j$s are $\geqslant$ or all are $\leqslant$.

Condition (ii) imposes natural restrictions on finiteness of rewards. Notice that we only require finiteness for adversaries which satisfy the probabilistic predicates contained in $\boldsymbol{\psi}$. We adopt this approach from [11] where, in addition, algorithms are given to check that $\boldsymbol{\psi}$ is reward-finite and to construct a modified MDP that is equivalent (in terms of satisfiability of $\boldsymbol{\psi}$) but for which $\boldsymbol{\psi}$ is *fully* reward-finite. This can be checked by a simpler multi-objective query containing only probabilistic predicates. Thus, in the remainder of this paper, we assume that all queries are fully reward-finite.

Condition (iii) ensures that the algorithms we define in the next section do not need to compute *unbounded expected total rewards* for MDPs with *both positive and negative rewards*, which is unsound. For unbounded reachability predicates, again using methods from [11], we can easily invert their bounds (to match those of any reward predicates) by making a simple change to the MDP.

**Extensions.** We also remark that the class of multi-objective properties outlined in this section can be extended in several respects. In particular, as shown in [11], we can add support for probabilistic $\omega$-regular (e.g. LTL) properties via reduction to probabilistic reachability on a product of the MDP and one or more deterministic Rabin automata. That work also allows arbitrary Boolean combinations of predicates, which are reduced to disjunctive normal form and treated separately. Both of these extensions can be adapted to our setting; the former we have implemented and used for our experiments in Section 5.

## 4 Multi-objective Probabilistic Model Checking

We now present efficient algorithms for checking the multi-objective queries defined in the previous section. Proofs of correctness can be found in the Appendix.

```
Input: MDP M, multi-objective predicate ψ = ([ρ_1]^{≤k_1}_{≥r_1}, ..., [ρ_n]^{≤k_n}_{≥r_n})
Output: true if ψ is achievable, false if not
X := ∅; ρ = (ρ_1, ... ρ_n); k = (k_1, ... k_n); r = (r_1, ... r_n);
do
    Find w separating r from down(X);
    Find adversary σ maximising ExpTot^{σ,k}_M(w·ρ);
    q := (ExpTot^{σ,k_i}_M(ρ_i))_{1≤i≤n};
    if w · q < w · r then return false;
    X := X ∪ {q};
while r ∉ down(X);
return true;
```

**Alg. 1.** Basic algorithm for checking achievability queries

**Reduction to basic form.** The first step when checking any type of query is to reduce the problem to one over a *basic* predicate on a modified MDP. We do so by converting reachability predicates into reward predicates (by adding a one-off reward of 1 upon reaching the target) and then negating objectives for predicates with upper bounds. Formally, we do the following.

**Proposition 2.** *Let* $\mathcal{M}=(S,\bar{s},\alpha,\delta)$ *be an MDP and* $\boldsymbol{\psi}=([T_1]^{\leqslant k_1}_{\sim_1 p_1},\ldots,[T_n]^{\leqslant k_n}_{\sim_n p_n},[\rho_1]^{\leqslant l_1}_{\bowtie_1 r_1},\ldots,[\rho_m]^{\leqslant l_m}_{\bowtie_m r_m})$ *be a multi-objective predicate. Let* $\mathcal{M}'=(S',(\bar{s},\emptyset),\alpha',\delta')$ *be the MDP defined as follows:* $S'=S\times 2^{\{1,\ldots,n\}}$, $\alpha'=\alpha\times 2^{\{1,\ldots,n\}}$ *and, for all* $s,s'\in S$, $a\in\alpha$ *and* $c\subseteq\{1,\ldots,n\}$*:*

- $\delta'((s,c),(a,c'))((s',c\cup c'))=\delta(s,a)(s')$ *where* $c'=\{i\mid s\in T_i\}\setminus c$*;*
- $\delta'((s,c),a')((s',c'))=0$ *for all other* $c$*,* $c'$ *and* $a'$*.*

*Now, let* $\boldsymbol{\psi}'$ *be* $([\rho_{T_1}]^{\leqslant k_1+1}_{\geqslant p_1},\ldots,[\rho_{T_n}]^{\leqslant k_n+1}_{\geqslant p_n},[\bar{\rho}_1]^{\leqslant l_1}_{\geqslant r_1},\ldots,[\bar{\rho}_m]^{\leqslant l_m}_{\geqslant r_m})$*, where: reward* $\rho_{T_i}((s,c),(a,c'))$ *is equal to* 1 *if* $i\in c'$ *and* $\sim_i\,=\,\geqslant$*, to* $-1$ *if* $\sim_i\,=\,\leqslant$*, and to* 0 *otherwise; and* $\bar{\rho}_i((s,c),(a,c'))$ *is equal to* $\rho_i(s,a)$ *if* $\bowtie_i\,=\,\geqslant$ *and to* $-\rho_i(s,a)$ *if* $\sim_i\,=\,\leqslant$*. Then* $\boldsymbol{\psi}$ *is satisfiable in* $\mathcal{M}$ *if and only if* $\boldsymbol{\psi}'$ *is satisfiable in* $\mathcal{M}'$*.*

Notice that the reduction described above results in reward structures with both positive and negative rewards. For time-bounded properties, this is not a concern. For unbounded ones, we must take care that they are all either non-negative or non-positive, as mentioned earlier in our discussion of condition (iii).

**Achievability queries.** We begin with achievability queries. We first give an outline of the overall algorithm; subsequently, we will describe in more detail how it is implemented in practice using value iteration.

By applying the reduction described above, we only need to consider the case of a *basic* multi-objective predicate $\boldsymbol{\psi}=([\rho_1]^{\leqslant k_1}_{\geqslant r_1},\ldots,[\rho_n]^{\leqslant k_n}_{\geqslant r_n})$. Alg. 1 shows how to check if $\boldsymbol{\psi}$ is satisfiable. It works by generating a sequence of weight vectors $\boldsymbol{w}$ and optimising a $\boldsymbol{w}$-weighted sum of the $n$ objectives. A resulting optimal adversary $\sigma$ is then used to generate a point $\boldsymbol{q}$ which is guaranteed to be contained on the Pareto curve for $\boldsymbol{\psi}$, and a collection $X$ of such points is assembled. Each new weight vector $\boldsymbol{w}$ is identified by finding a separating hyperplane between $down(X)$ and $\boldsymbol{r}=(r_1,\ldots,r_n)$. Once $\boldsymbol{r}$ is found to be contained in $down(X)$, we know that $\boldsymbol{\psi}$ is achievable. If, on the other hand, $\boldsymbol{w} \cdot \boldsymbol{q} < \boldsymbol{w} \cdot \boldsymbol{r}$, then we know that $\boldsymbol{\psi}$ cannot possibly be achievable.

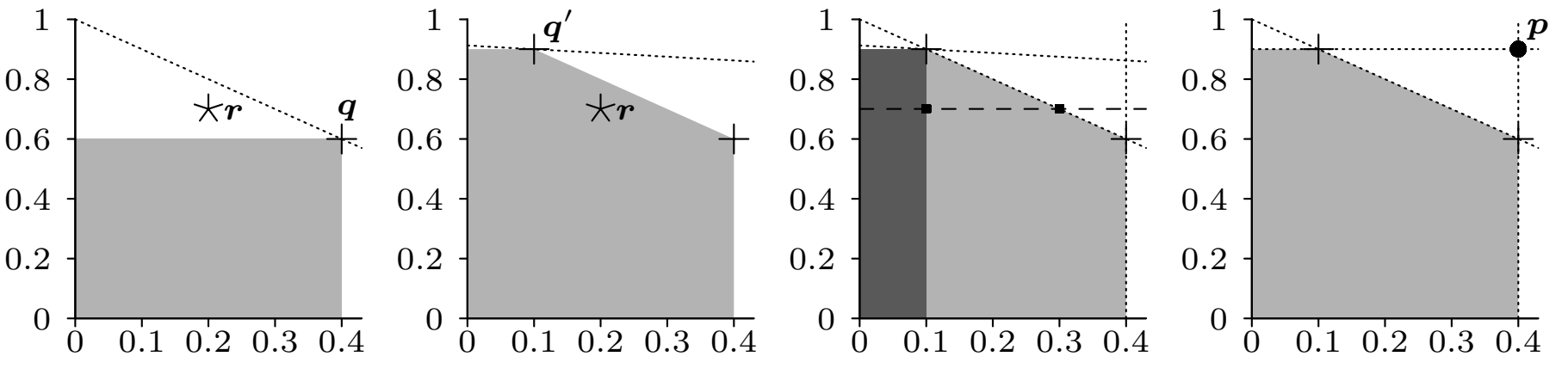

**Fig. 2.** Example executions of Algorithms 1, 3 and 4 (see Examples 2, 3 and 4).

Correctness of Alg. 1 is proved in Appx. A.2. Termination is guaranteed by the fact that each iteration of the loop identifies a point $\boldsymbol{q}$ on a unique face of the Pareto curve. In the worst case, the number of faces is exponential in $|\mathcal{M}|$, $k$ and $n$ [9]; however, our experimental results (in Section 5) show the number of steps is usually small on practical examples. The individual model checking problems solved in each step (lines 4-5 of Alg. 1) require time polynomial in $|\mathcal{M}|$. We describe their practical implementation in the next section.

**Example 2.** We illustrate the execution of Alg. 1 on the MDP from Example 1 and the achievability query $([\{s_1\}]_{\geqslant 0.2}, [\{s_3\}]_{\geqslant 0.7})$. Let us assume that we have already applied Proposition 2 so that we have an equivalent reward predicate $([\rho_1]_{\geqslant 0.2}, [\rho_2]_{\geqslant 0.7})$ (the full reduction is in Appx. A.1). As a first (arbitrary) weight vector, we pick $\boldsymbol{w}{=}(0.5, 0.5)$ and then maximise $\boldsymbol{w} \cdot \boldsymbol{\rho}$. The resulting optimal adversary $\sigma$ (which chooses $a$ in $s_0$) gives $\boldsymbol{q} = (\mathit{ExpTot}^{\sigma}_{\mathcal{M}}(\rho_1), \mathit{ExpTot}^{\sigma}_{\mathcal{M}}(\rho_2))$ $= (0.4, 0.6)$ and we have $X = \{\boldsymbol{q}\}$. Fig. 2(a) shows the point $\boldsymbol{q}$ (as $+$) and the target point $\boldsymbol{r} = (0.2, 0.7)$ (as $\star$). The dotted line represents the hyperplane with orientation $\boldsymbol{w}$ passing through $\boldsymbol{q}$ (i.e. the points $\boldsymbol{x}$ for which $\boldsymbol{w} \cdot \boldsymbol{x} = \boldsymbol{w} \cdot \boldsymbol{q} = 0.5$), the points above which correspond to unachievable value pairs. The grey region is $\mathit{down}(X)$, in which all points are achievable. Since $\boldsymbol{r} \notin \mathit{down}(X)$, we continue.

Next, we pick a weight $\boldsymbol{w}' = (0.1, 0.8)$. Maximising $\boldsymbol{w}' \cdot \boldsymbol{\rho}$ results in adversary $\sigma'$ (which chooses $b$ in $s_0$) giving $\boldsymbol{q}' = (0.1, 0.9)$, which we add to $X$. Fig. 2(b) again shows both points in $X$, $\mathit{down}(X)$ and $\boldsymbol{r}$. It also plots points $\boldsymbol{x}$ for which $\boldsymbol{w}' \cdot \boldsymbol{x} = \boldsymbol{w}' \cdot \boldsymbol{q}' = 0.73$. Since $\boldsymbol{r}$ is now in $\mathit{down}(X)$, the algorithm returns true.

**Value iteration.** The most expensive part of Alg. 1 in practice is the combination of lines 4-5, which computes the maximum possible value for a weighted sum of reward objectives, determines a corresponding optimal adversary $\sigma$, and then finds the value for the $n$ individual objectives under $\sigma$.

Alg. 2 shows how to perform all these tasks using a value iteration-style computation. One key difference between this algorithm and standard value iteration is that it needs to optimise a combination of unbounded and bounded properties. This is done in three phases (lines 3-8, 9-13 and 14-20). The first two correspond to the unbounded part; the third to the bounded part.

Another important difference is that the algorithm performs the optimisation of the weighted sum $\boldsymbol{w}{\cdot}\boldsymbol{\rho}[\boldsymbol{k}]$ and the computation of the vector of individual objective values $\boldsymbol{q} = (\mathit{ExpTot}^{\sigma,k_i}_{\mathcal{M}}(\rho_i))_{1 \leqslant i \leqslant n}$ simultaneously: the former in phases

**Input**: MDP $\mathcal{M}=(S,\bar{s},\alpha,\delta)$, weight vect. $\boldsymbol{w}$, reward structures $\boldsymbol{\rho}=(\rho_1,\ldots,\rho_n)$, vector of time bounds $\boldsymbol{k}\in(\mathbb{N}\cup\{\infty\})^n$, convergence threshold $\varepsilon$
**Output**: Adv. $\sigma$ maximising $\mathit{ExpTot}^{\sigma,\boldsymbol{k}}_{\mathcal{M}}(\boldsymbol{w}\cdot\boldsymbol{\rho})$, $\boldsymbol{q}=(\mathit{ExpTot}^{\sigma,k_i}_{\mathcal{M}}(\rho_i))_{1\leqslant i\leqslant n}$

1 $\boldsymbol{x}:=\boldsymbol{0};\ \boldsymbol{x}^1:=\boldsymbol{0};\ \ldots;\ \boldsymbol{x}^n:=\boldsymbol{0};\ \boldsymbol{y}=\boldsymbol{0};\ \boldsymbol{y}^1:=\boldsymbol{0};\ \ldots;\ \boldsymbol{y}^n:=\boldsymbol{0};$
2 $\sigma^\infty(s)=\bot$ for all $s\in S$;
3 **do**
4   **foreach** $s\in S$ **do**
5     $y_s:=\max_{a\in A(s)}(\sum_{\{i|k_i=\infty\}} w_i\cdot\rho_i(s,a)+\sum_{s'\in S}\delta(s,a)(s')\cdot x_{s'});$
6     $\sigma^\infty(s):=\arg\max_{a\in A(s)}(\sum_{\{i|k_i=\infty\}} w_i\cdot\rho_i(s,a)+\sum_{s'\in S}\delta(s,a)(s')\cdot x_{s'});$
7   $\delta:=\max_{s\in S}(y_s-x_s);\ \boldsymbol{x}:=\boldsymbol{y};$
8 **while** $\delta>\varepsilon$;
9 **do**
10   **foreach** $s\in S$ *and* $i\in\{1,\ldots,n\}$ *where* $k_i=\infty$ **do**
11     $y^i_s:=\rho_i(s,\sigma^\infty(s))+\sum_{s'\in S}\delta(s,\sigma^\infty(s))(s')\cdot x^i_{s'};$
12   $\delta:=\max_{i=1}^n\max_{s\in S}(y^i_s-x^i_s);\ \boldsymbol{x}^1:=\boldsymbol{y}^1;\ \ldots;\ \boldsymbol{x}^n:=\boldsymbol{y}^n;$
13 **while** $\delta>\varepsilon$;
14 **for** $j=\max\{k_\ell<\infty\mid\ell\in\{1,\ldots,n\}\}$ *down to* 1 **do**
15   **foreach** $s\in S$ **do**
16     $y_s:=\max_{a\in A(s)}(\sum_{\{i|k_i\geqslant j\}} w_i\cdot\rho_i(s,a)+\sum_{s'\in S}\delta(s,a)(s')\cdot x_{s'});$
17     $\sigma^j(s):=\arg\max_{a\in A(s)}(\sum_{\{i|k_i\geqslant j\}} w_i\cdot\rho_i(s,a)+\sum_{s'\in S}\delta(s,a)(s')\cdot x_{s'});$
18     **foreach** $i\in\{1,\ldots,n\}$ *where* $k_i\geqslant j$ **do**
19       $y^i_s:=\rho_i(s,\sigma^j(s))+\sum_{s'\in S}\delta(s,\sigma^j(s))(s')\cdot x^i_{s'};$
20   $\boldsymbol{x}:=\boldsymbol{y};\ \boldsymbol{x}^1:=\boldsymbol{y}^1;\ \ldots;\ \boldsymbol{x}^n:=\boldsymbol{y}^n;$
21 **foreach** $i\in\{1,\ldots,n\}$ **do** $q_i:=y^i_{\bar{s}};$
22 $\sigma$ behaves as $\sigma^j$ in $j$-th step when $j<\max_{i\in\{1,\ldots,n\}}k_i$, and as $\sigma^\infty$ afterwards;
23 **return** $\sigma,\boldsymbol{q}$

**Alg. 2.** Value iteration-based algorithm for lines 4-5 of Alg. 1.

1 and 3; the latter in phases 2 and 3. Consider first the optimisation of $\boldsymbol{w}\cdot\boldsymbol{\rho}[\boldsymbol{k}]$. The values, for all states $s\in S$, are computed as a sequence of increasingly precise approximations, stored in a pair of vectors, $\boldsymbol{x}$ and $\boldsymbol{y}$. Each new approximation is stored in $\boldsymbol{y}$ (line 5); then, $\boldsymbol{x}$ and $\boldsymbol{y}$ are compared for convergence and $\boldsymbol{x}$ is set to $\boldsymbol{y}$ (line 7) before proceeding to the next iteration. Computation of the bounded part of $\boldsymbol{w}\cdot\boldsymbol{\rho}[\boldsymbol{k}]$ continues in phase 3 in similar fashion (although no convergence check is needed). During optimisation of $\boldsymbol{w}\cdot\boldsymbol{\rho}[\boldsymbol{k}]$, a corresponding optimal adversary is also determined, with the unbounded and bounded fragments stored in $\sigma^\infty$ and $\sigma^j$, respectively. The choices made by this adversary are used to compute the value $q_i$ for each of the $n$ individual objectives, the values for which are also stored in pairs of vectors ($\boldsymbol{x}^i,\boldsymbol{y}^i$ for each $q_i$).

In practice, storing multiple $|S|$-sized vectors ($\boldsymbol{x}$, $\boldsymbol{y}$, $\boldsymbol{x}^i$, and $\boldsymbol{y}^i$) is relatively expensive. We discuss later how the algorithm's memory usage can be improved. We include an example of the execution of Alg. 2 in Appx. A.1.

**Numerical queries.** We now turn our attention to numerical queries. Alg. 3 shows how Alg. 1 can be adapted to check these. Like Alg. 1, it generates points $\boldsymbol{q}$ on the Pareto curve from a sequence of weight functions $\boldsymbol{w}$. For the objective $\rho_1$ that is being optimised, we generate a sequence of lower bounds $r_1$ that are **Input**: MDP $\mathcal{M}$, objective $[\rho_1]^{\leqslant k_1}_{\max}$, predicate $([\rho_2]^{\leqslant k_2}_{\geqslant r_2}, \ldots, [\rho_n]^{\leqslant k_n}_{\geqslant r_n})$

**Output**: Value of $num([\rho_1]^{\leqslant k_1}_{\max}, ([\rho_2]^{\leqslant k_2}_{\geqslant r_2}, \ldots, [\rho_n]^{\leqslant k_n}_{\geqslant r_n}))$

1 $X{:=}\emptyset$; $\boldsymbol{\rho}{:=}(\rho_1, \ldots \rho_n)$; $\boldsymbol{k}{:=}(k_1, \ldots k_n)$; $\boldsymbol{r}{:=}(\min_{\sigma \in Adv_M} ExpTot^{\sigma,k_1}_{\mathcal{M}}(\rho_1), r_2, \ldots r_n)$;

2 **do**

3     Find $\boldsymbol{w}$ separating $\boldsymbol{r}$ from $down(X)$ such that $w_1 > 0$;

4     Find adversary $\sigma$ maximising $ExpTot^{\sigma,\boldsymbol{k}}_{\mathcal{M}}(\boldsymbol{w}{\cdot}\boldsymbol{\rho})$;

5     $\boldsymbol{q} := (ExpTot^{\sigma,k_i}_{\mathcal{M}}(\rho_i))_{1 \leqslant i \leqslant n}$;

6     **if** $\boldsymbol{w} \cdot \boldsymbol{q} < \boldsymbol{w} \cdot \boldsymbol{r}$ **then return** $\bot$;

7     $X := X \cup \{\boldsymbol{q}\}$; $\quad r_1 := \max\{r_1, \max\{r' \mid (r', r_2, \ldots, r_n) \in down(X)\}\}$;

8 **while** $\boldsymbol{r} \notin down(X)$ *or* $\boldsymbol{w} \cdot \boldsymbol{q} > \boldsymbol{w} \cdot \boldsymbol{r}$;

9 **return** $r_1$;

**Alg. 3.** Algorithm for checking numerical queries used in the same fashion as Alg. 1. Initially, we take $r_1$ to be the minimum possible value for $\rho_1$, which can be computed with a separate instance of value iteration. New (non-decreasing) values for $r_1$ are generated at each step based on the set of points $X$ determined so far. The numerical computation for each step (lines 4-5 of Alg. 3) can again be carried out with Alg. 2. Correctness of Alg. 3 is proved in Appx. A.4. The bound on the number of steps needed is as for Alg. 1.

**Example 3.** We demonstrate Alg. 3 on the MDP from Example 1 and numerical query $([\{s_1\}]_{\max}, ([\{s_3\}]_{\geqslant 0.7}))$. Initially, $r_1{=}0.1$ and, with $\boldsymbol{w}{=}(0.1, 0.8)$, we get $\boldsymbol{q}{=}(0.1, 0.9)$. The resulting area $down(X)$ is shown as dark grey in Fig. 2(c). Next, $r_1$ remains as 0.1 and, with $\boldsymbol{w}{=}(1, 0)$, we get $\boldsymbol{q}{=}(0.4, 0.6)$. Adding this to $X$, $down(X)$ is enlarged by the light grey area. Finally, $r_1$ is set to 0.3, choosing $\boldsymbol{w}{=}(0.5, 0.5)$ yields $\boldsymbol{q}{=}(0.4, 0.6)$ again, and the loop ends. Fig. 2(c) also shows the points $\boldsymbol{q}$ and $\boldsymbol{r}$ (as + and ■). The final value returned is $r_1{=}0.3$.

**Pareto curves.** Next, we discuss Pareto queries. Generating and visualising Pareto curves (or their approximations) provides a much clearer view of the trade-offs between objectives. Our algorithm is implemented as a simple modification of our previous algorithms, and is presented as Alg. 4. For simplicity, we focus on the 2-objective case, which is most practical for visualisation. Our implementation, described later, also supports the 3-objective case and, in theory, this can be extended to an arbitrary number of objectives.

Alg. 4, like the earlier ones, builds a set $X$ of points on a Pareto curve $P$ using weights $\boldsymbol{w}$. Since $P$ is convex, the surface of points $X$ represents a *lower* approximation of $P$. Our algorithm also constructs an *upper* approximation $Y$ using the generated weights $\boldsymbol{w}$. As illustrated in Example 2, for each point $\boldsymbol{q} \in X$, there is a corresponding hyperplane passing through $\boldsymbol{q}$ and with orientation $\boldsymbol{w}$, above which no values are achievable. Hence these represent upper bounds on $P$ and we store, in $Y$, any weight $\boldsymbol{w}$ that resulted in each point $\boldsymbol{q} \in X$.

The sequence of weights $\boldsymbol{w}$ is generated as follows. We construct an initial curve using weights $(1, 0)$ and $(0, 1)$. Then, we repeatedly: (i) sort the points in $X$; (ii) for each successive pair $\boldsymbol{x}^i, \boldsymbol{x}^{i+1}$ in $X$, find the lowest point $\boldsymbol{p}$ on the intersection of the hyperplanes stored in $Y$ for $\boldsymbol{x}^i$ and $\boldsymbol{x}^{i+1}$; (iii) choose $\boldsymbol{w}$

**Input**: MDP $\mathcal{M}$, reward structures $\boldsymbol{\rho} = (\rho_1, \rho_2)$, time bounds $(k_1, k_2)$, $\varepsilon_p \in \mathbb{R}_{>0}$
**Output**: An $\varepsilon_p$-approximation of a Pareto curve

$X := \emptyset$; $Y : \mathbb{R}^2 \to 2^{\mathbb{R}^2}$, initially $Y(x) = \emptyset$ for all $x$; $\boldsymbol{w} = (1, 0)$;
Find adversary $\sigma$ maximising $ExpTot^{\sigma,\boldsymbol{k}}_{\mathcal{M}}(\boldsymbol{w}\cdot\boldsymbol{\rho})$;
$\boldsymbol{q} := (ExpTot^{\sigma,k_1}_{\mathcal{M}}(\rho_1), ExpTot^{\sigma,k_2}_{\mathcal{M}}(\rho_2))$;
$X := X \cup \{\boldsymbol{q}\}$; $Y(\boldsymbol{q}) := Y(\boldsymbol{q}) \cup \{\boldsymbol{w}\}$; $\boldsymbol{w} = (0, 1)$;
**do**
  Find adversary $\sigma$ maximising $ExpTot^{\sigma,\boldsymbol{k}}_{\mathcal{M}}(\boldsymbol{w}\cdot\boldsymbol{\rho})$;
  $\boldsymbol{q} := (ExpTot^{\sigma,k_1}_{\mathcal{M}}(\rho_1), ExpTot^{\sigma,k_2}_{\mathcal{M}}(\rho_2))$;
  $X := X \cup \{\boldsymbol{q}\}$; $Y(\boldsymbol{q}) := Y(\boldsymbol{q}) \cup \{\boldsymbol{w}\}$; $\boldsymbol{w} = \bot$;
  Order $X$ to a sequence $\boldsymbol{x}^1, \ldots, \boldsymbol{x}^m$ such that $\forall\ i$: $x^i_1 \leqslant x^{i+1}_1$ and $x^i_2 \geqslant x^{i+1}_2$;
  **foreach** $i \in \{1, \ldots m-1\}$ **do**
    Let $\boldsymbol{u}$ be the element of $Y(\boldsymbol{x}^i)$ with maximal $u_1$;
    Let $\boldsymbol{u'}$ be the element of $Y(\boldsymbol{x}^{i+1})$ with minimal $u'_1$.;
    Find a point $\boldsymbol{p}$ such that $\boldsymbol{u} \cdot \boldsymbol{p} = \boldsymbol{u} \cdot \boldsymbol{x}^i$ and $\boldsymbol{u'} \cdot \boldsymbol{p} = \boldsymbol{u'} \cdot \boldsymbol{x}^{i+1}$;
    **if** *distance of* $\boldsymbol{p}$ *from* $down(X)$ *is* $\geqslant \varepsilon_p$ **then**
      Find $\boldsymbol{w}$ separating $down(X)$ from $\boldsymbol{p}$, maximising $\boldsymbol{w}\cdot\boldsymbol{p} - \max_{\boldsymbol{x} \in down(X)} \boldsymbol{w}\cdot\boldsymbol{x}$;
**while** $\boldsymbol{w} \neq \bot$;
**return** $X$

**Alg. 4.** Algorithm for Pareto curve approximation for 2 objectives as a separating hyperplane between $down(X)$ and $\boldsymbol{p}$. The algorithm continues until the maximum distance between the two approximations falls below some threshold $\varepsilon_p$. In principal, the algorithm can enumerate all faces of $P$. The reason for constructing an $\varepsilon_p$-approximation is two-fold: firstly, the number of faces is potentially large, whereas an approximation may suffice; secondly, computation of individual points (using value iteration), is already approximate.

**Example 4.** We illustrate Alg. 4 on the MDP from Example 1 with objectives $([\{s_1\}]_{\max}, [\{s_3\}]_{\max})$. The first two weight vectors $\boldsymbol{w}$ are $(1, 0)$ and $(0, 1)$, yielding points $\boldsymbol{q}$ of $(0.4, 0.6)$ and $(0.1, 0.9)$, repectively (see Fig. 2(d)). The hyperplanes attached to each point are also shown, by dotted lines, as is their intersection $\boldsymbol{p} = (0.4, 0.9)$. We choose separating hyperplane $\boldsymbol{w} = (0.5, 0.5)$, indicated by the sloped dotted line. The algorithm then finds the intersection $(0.1, 0.9)$ of this with the horizontal line and, since this point is already in $down(X)$, terminates.

**Adversary generation.** Finally, we describe how to generate *optimal adversaries* for our multi-objective queries. We explain this for achievability queries, but it can easily be adapted to the other types too. Unlike standard (single-objective) MDP model checking, where deterministic adversaries always suffice to optimise reachability/reward objectives, multi-objective optimisation requires randomised adversaries. Alg. 1, when finding that bounds $\boldsymbol{r}$ are achievable, generates points $\boldsymbol{q}^1, \ldots, \boldsymbol{q}^m$ on the Pareto curve. Each corresponding call to Alg. 2 returns a (deterministic) adversary, say $\sigma_{\boldsymbol{q}^j}$ for the current point $\boldsymbol{q}^j$. The final adversary $\sigma_{\mathrm{opt}}$ is constructed from these and a weight vector $\boldsymbol{u} \in \mathbb{R}^m$ satisfying $r_i \leqslant \sum_{j=1}^{m} u_i \cdot q^j_i$ for all $1 \leqslant i \leqslant n$: it simply makes an initial one-off random choice of adversary $\sigma_{\boldsymbol{q}^j}$ to mimic (each with probability $u_j$).

## 5 Implementation and Results

We implemented our multi-objective model checking techniques in PRISM [14], also adding the automaton construction of [11] to support $\omega$-regular properties. Value iteration is built on top of PRISM's "sparse" engine. It would also be straightforward to adapt its symbolic (MTBDD) engine, which can improve scalability on models exhibiting regularity; but, for the current set of experiments, the sparse engine suffices to illustrate the benefits offered by our approach.

**Heuristics and optimisations.** Our core algorithms are based on generating weight vectors $\boldsymbol{w}$ representing separating hyperplanes (e.g. at line 3 of Alg. 1). The choice of each $\boldsymbol{w}$ is not unique and affects the number of steps needed by the algorithm. Based on our results, the following is an effective heuristic. For the first $n$ vectors (assuming $n$ objectives), choose $\boldsymbol{w}$ with $w_i = 1$ for some $i$. Next, given point $\boldsymbol{r}$ and set of points $X$, choose $\boldsymbol{w}$ to maximise $\min_{\boldsymbol{x} \in X}(\boldsymbol{w} \cdot \boldsymbol{q} - \boldsymbol{w} \cdot \boldsymbol{x})$, i.e., pick the hyperplane with maximal Euclidean distance $d$ from $\boldsymbol{q}$. This is done by solving the LP: "maximise $d$ subject to $\sum_{i=1}^{n} w_i = 1$ and $w_i \cdot (q_i - x_i) \geqslant d$ for all $\boldsymbol{x} \in X$". In practice, these problems are small and fast to solve.

We also apply various optimisations to the basic value iteration algorithms of Section 4. For unbounded properties, Gauss-Seidel value iteration [19] can be used to increase performance. Furthermore, we can significantly reduce the number of vectors stored with slight changes to Alg. 2; details are in Appx. A.6.

**Experimental results.** We evaluated our techniques on benchmarks from several sources.[2] First, we used multi-objective problems resulting from the assume-guarantee framework of [15]. Second, we verified multi-objective properties on existing PRISM models: (i) a task-graph *scheduler* problem, minimising expected job completion time and expected energy consumption; (ii) a *team-formation* protocol, maximising the probability of completing two (separate) tasks and the expected size of the team that does so; (iii) a dynamic power management (*dpm*) controller, minimising over $k$ steps both expected energy consumption and expected average queue size. Experiments were run on a 2.66GHz PC with 8GB of RAM. We used $\varepsilon = 10^{-6}$ for value iteration (this is the default in PRISM; smaller values led to very similar results) and $\varepsilon_p = 10^{-4}$ for Pareto curve generation.

The results are shown in Table 1: assume-guarantee problems at the top; the others below. For each model, we give the size (number of states), and details of the objectives in the query used. The middle part of the table compares the performance of our value iteration-based technique with the LP-based implementation of [11] on numerical queries. In our experiments, performance for achievability queries was very similar, so we omit them. The right part of the table shows times to compute Pareto curves for the same objectives on each model (which cannot be done with the implementation of [11]). For value iteration-based algorithms, we show the number of points (steps of the algorithm) needed; for LP-based, we show the size of the linear program solved.

Comparing the value-iteration and LP-based approaches, we see huge gains in run-time for our methods (up to approx. 150 times faster). There are also significant improvements in scalability: the biggest models solved with value iteration are about 20 times bigger than those for LP. One factor in the low run-times for our technique is that the algorithms generally require a fairly small number of steps, even when generating the Pareto curve.

[2] All models/properties used are at www.prismmodelchecker.org/files/atva12mo/.

| Case study [parameters] | | Num. states | Objectives Num. | Objectives Types | Numerical query LP ([11]) LP size | Numerical query LP ([11]) Time (s) | Numerical query Val. iter. Pt.s | Numerical query Val. iter. Time (s) | Pareto query Val. iter. Pt.s | Pareto query Val. iter. Time (s) |
|---|---|---|---|---|---|---|---|---|---|---|
| *consensus* | 3 2 | 691 | | $[T_1]_{\max}$ | 1026 | 0.57 | 3 | **0.02** | 3 | 0.04 |
| *(2 proc.s)* | 4 2 | 1517 | 2 | | 2288 | 0.67 | 3 | **0.03** | 3 | 0.05 |
| $[R\ K]$ | 5 2 | 3169 | | $[T_2]_{\max}$ | 4812 | 0.94 | 3 | **0.05** | 3 | 0.06 |
| *consensus* | 3 2 | 17455 | | $[T_1]_{\max}$ | 40386 | 9.85 | 3 | **0.22** | 3 | 0.27 |
| *(3 proc.s)* | 4 2 | 61017 | 2 | | 140676 | 144.06 | 3 | **0.87** | 3 | 1.06 |
| $[R\ K]$ | 5 2 | 181129 | | $[T_2]_{\max}$ | *mem-out* | | 3 | **2.83** | 3 | 3.44 |
| *zeroconf* | 4 | 5449 | | $[T_1]_{\max}$ | 12916 | 1.25 | 2 | **0.13** | 4 | 0.60 |
| $[K]$ | 6 | 10543 | 2 | | 24639 | 7.07 | 4 | **0.46** | 4 | 0.79 |
| | 8 | 17221 | | $[T_2]_{\max}$ | 40833 | 19.6 | 4 | **0.76** | 4 | 1.13 |
| *zeroconf-tb* | 2 14 | 29572 | | $[T_1]_{\max}$ | 61816 | 5.25 | 3 | **1.69** | 2 | 0.85 |
| $[K\ T]$ | 4 10 | 19670 | 2 | | 46659 | 5.01 | 2 | **0.32** | 3 | 0.84 |
| | 4 14 | 42968 | | $[T_2]_{\max}$ | 103964 | 11.01 | 2 | **0.63** | 3 | 1.77 |
| *team-form.* | 3 | 12475 | | $[T_1]_{\max}$ | 14935 | 1.37 | 4 | **0.21** | 7 | 0.24 |
| $[N]$ | 4 | 96665 | 2 | | 115289 | 11.57 | 4 | **1.08** | 7 | 1.72 |
| | 5 | 907993 | | $[\rho_2]_{\max}$ | *mem-out* | | 2 | **5.66** | 6 | 12.66 |
| *team-form.* | 3 | 12475 | | $[T_1]_{\max}$ | 14935 | 1.37 | 3 | **0.18** | 57 | 1.39 |
| $[N]$ | 4 | 96665 | 3 | $[T_2]_{\max}$ | 115289 | 10.55 | 5 | **1.77** | 61 | 14.55 |
| | 5 | 907993 | | $[\rho_2]_{\max}$ | *mem-out* | | 2 | **9.49** | 57 | 141.76 |
| *scheduler* | 5 | 31965 | | $[\rho_1]_{\min}$ | 57954 | 59.15 | 8 | **6.10** | 10 | 8.08 |
| $[K]$ | 25 | 633735 | 2 | | *mem-out* | | 8 | **526.56** | 11 | 776.44 |
| | 50 | 2457510 | | $[\rho_2]_{\min}$ | *mem-out* | | 8 | **3938.94** | 10 | 5361.86 |
| *dpm* | 100 | 636 | | $[\rho_1]^{\leqslant k}_{\min}$ | n/a | n/a | 3 | **4.50** | 6 | 0.12 |
| $[k]$ | 200 | 636 | 2 | | n/a | n/a | 3 | **4.30** | 11 | 0.32 |
| | 300 | 636 | | $[\rho_2]^{\leqslant k}_{\min}$ | n/a | n/a | 3 | **4.59** | 9 | 0.36 |

**Table 1.** Experimental results for our implementation and a comparison with [11].

**Pareto curves.** Finally, we show in Fig. 3 the Pareto curves generated for some of our examples. Plot (a) shows, for a task-graph scheduling problem, how different schedulers vary in terms of completion time and energy usage. Plot (b) is from an instance of assume-guarantee verification; the plot shows how it is possible to bound the probability of an error in the overall system (y-axis) for various different reliability levels of one of the components (x-axis). Plot (c) shows a 3-objective Pareto curve evaluating strategies in a team-formation protocol (see Fig. 3 caption for objectives). In each case, the plots give a clear, visual illustration of the trade-off between competing objectives. The curves could also be used to quickly answer any additional achievability or numerical queries for those objectives, without running any further model checking.

## 6 Conclusions

We have presented novel techniques for multi-objective model checking of MDPs, using a value iteration-based computation to build successive approximations of the Pareto curve. Compared to existing approaches, this gives significant gains in efficiency and scalability, and enables verification of time-bounded properties. Furthermore, we showed the benefits of visualising the Pareto curve for several probabilistic model checking case studies. Future directions include extending our techniques to timed probabilistic models such as CTMDPs and PTAs.

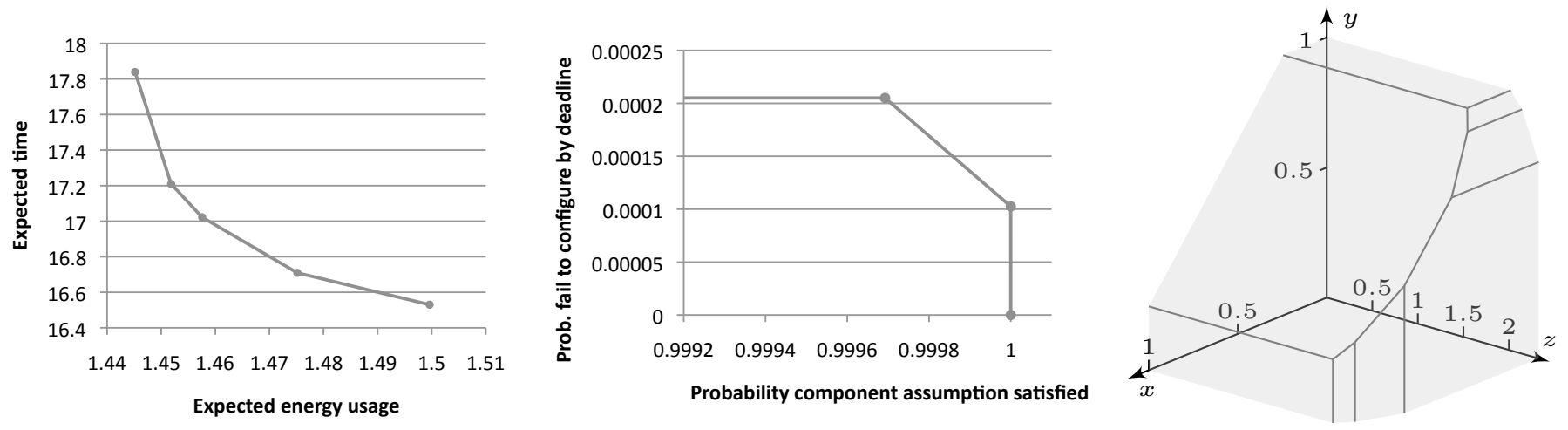

**Fig. 3.** Pareto curves from: (a) task-graph scheduler, $K{=}2$; (b) Zeroconf protocol, $K{=}2, T{=}10$; (c) team formation protocol, $N{=}3$ (axes x/y/z = Probability of completing task 1/probability of completing task 2/expected size of successful team)

**Acknowledgments.** The authors are part supported by ERC Advanced Grant VERIWARE and EPSRC grant EP/F001096/1. Vojtěch Forejt is also supported by a Royal Society Newton Fellowship.

# A Appendix

This appendix contains additional details for the paper's running example (A.1), proofs omitted from the main text (A.2–A.5) and optimisation details for our implementation (A.6).

## A.1 Additional Details for the Running Example

**Reduction for Example 2.** Example 2 describes the execution of Alg. 1 on the MDP $\mathcal{M}$ of Fig. 1(a) and achievability query $\boldsymbol{\psi} = ([\{s_1\}]_{\geqslant 0.2}, [\{s_3\}]_{\geqslant 0.7})$. We describe here the application of Proposition 2, which converts to a query in basic form on a modified MDP $\mathcal{M}'$. We give (the reachable fragment of) $\mathcal{M}'$ in Fig. 4. We also show three reward structures, $\bar{\rho}$, $\rho_{\{s_1\}}$ and $\rho_{\{s_2\}}$; the second two are for this example, the third is used below. The equivalent query, in basic form, is now $\boldsymbol{\psi}' = ([\rho_{\{s_1\}}]_{\geqslant 0.2}, [\rho_{\{s_3\}}]_{\geqslant 0.7})$. Note that, in Example 2, we did not use $\mathcal{M}'$ for clarity of presentation. To make the example formally correct, we would have to replace each $\mathcal{M}$ and $s_0$ with $\mathcal{M}'$ and $(s_0, \emptyset)$.

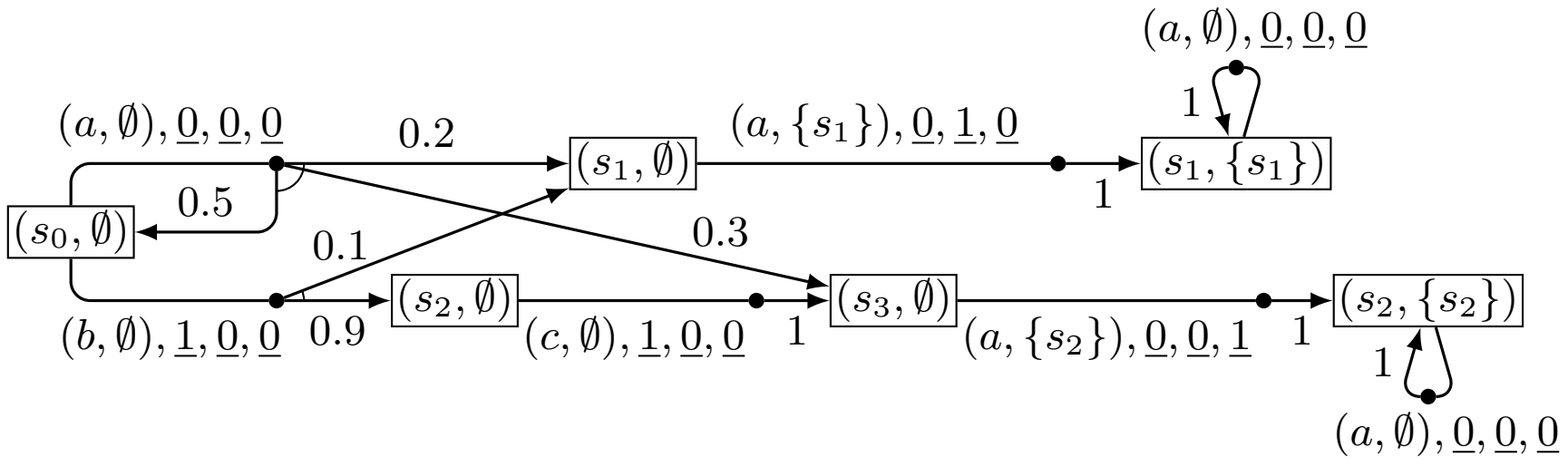

**Fig. 4.** The MDP $\mathcal{M}'$ from Example 2.

We also include an additional example illustrating the value iteration algorithm.

**Example 5.** We explain the flow of Alg. 2 on the MDP $\mathcal{M}'$ (from above) and the multi-objective query $([\rho_{\{s_1\}}]^{\leqslant 3}_{\geqslant x}, [\bar{\rho}]_{\geqslant y})$. Suppose $\boldsymbol{w} = (0.3, 0.7)$. The first and second do-while loops in the algorithm will be performed twice, yielding the values $x_{(s_0,\emptyset)} = 1.33$, $x^1_{(s_0,\emptyset)} = 0$ and $x^2_{(s_0,\emptyset)} = 1.9$. Then, we proceed with the third do-while loop, which will be performed 3 times, giving the values $x_{(s_0,\emptyset)} = 1.36$, $x^1_{(s_0,\emptyset)} = 0.1$ and $x^2_{(s_0,\emptyset)} = 1.9$. The returned vector is $\boldsymbol{q} = (0.1, 1.9)$.

## A.2 Correctness of Alg. 1

To prove the correctness of Alg. 1, we need the two lemmas below. Recall that a set $Y$ is a convex polytope if it is a set of all convex combinations of some finite set $X$, and that a *face* of convex polytope $Y$ is a set $Y' \subseteq Y$ such that there is a point $\boldsymbol{v}$ with $\boldsymbol{y'} \cdot \boldsymbol{v} = \boldsymbol{y''} \cdot \boldsymbol{v}$ and $\boldsymbol{y'} \cdot \boldsymbol{v} > \boldsymbol{y} \cdot \boldsymbol{v}$ for all $\boldsymbol{y} \notin Y$ and $\boldsymbol{y'}, \boldsymbol{y''} \in Y'$.

**Lemma 1.** *Let $X$ be a convex polytope. For every $\boldsymbol{w}$ there is a face $Y$ such that $\boldsymbol{y'} \cdot \boldsymbol{w} = \boldsymbol{y''} \cdot \boldsymbol{w}$ and $\boldsymbol{y'} \cdot \boldsymbol{w} > \boldsymbol{y} \cdot \boldsymbol{w}$ for all $\boldsymbol{y} \notin Y$ and $\boldsymbol{y'}, \boldsymbol{y''} \in Y'$.*

*Proof.* It suffices to see that there is a point $x$ such that $\boldsymbol{x} \cdot \boldsymbol{w} \geqslant \boldsymbol{y} \cdot \boldsymbol{w}$ for all $y \in X$. This follows from the fact that a polyhedron is a closed set. Then, we take $Y = \{\boldsymbol{y} \mid \boldsymbol{x} \cdot \boldsymbol{w} = \boldsymbol{y} \cdot \boldsymbol{w}\}$, and we are finished by the definition of a face.

**Lemma 2.** *Let $\boldsymbol{\psi} = ([\rho_1]^{\leqslant k_1}_{\geqslant r_1}, \dots, [\rho_n]^{\leqslant k_n}_{\geqslant r_n})$ be a multi-objective predicate, $X \subseteq \mathbb{R}^n$ be a set of vectors $(r_1, \dots, r_n)$ such that $\boldsymbol{\psi}$ is satisfiable, $\boldsymbol{p} = (p_1, \dots p_n)$ be a point not in $down(X)$ and $\boldsymbol{w}$ be a weight vector separating $\boldsymbol{p}$ from $down(X)$. For an adversary $\sigma$ maximising $ExpTot^{\sigma}_{\mathcal{M}}(\boldsymbol{w}\cdot\boldsymbol{\rho}[\boldsymbol{k}])$ and $\boldsymbol{q} = (ExpTot^{\sigma}_{\mathcal{M}}(\rho_i))_{1 \leqslant i \leqslant n}$:*

*(i) If $\boldsymbol{w} \cdot \boldsymbol{q} < \boldsymbol{w} \cdot \boldsymbol{p}$, then $\boldsymbol{\psi}$ is not satisfiable.*
*(ii) If $\boldsymbol{w} \cdot \boldsymbol{q} \geqslant \boldsymbol{w} \cdot \boldsymbol{p}$, then there is a face $F$ (of the convex polytope of achievable solutions) such that $\boldsymbol{q}$ is on $F$, but no point from $X$ is in $F$.*

*Proof.* Let us start with the first item and prove it by contrapositive. If $\boldsymbol{\psi}$ is achievable, then there must be $\boldsymbol{p'}$ such that $p'_i \geqslant p_i$ for all $i$ and an adversary $\sigma'$ such that $ExpTot^{\sigma',k_i}_{\mathcal{M}}(\rho_i) = p'_i$ for all $i$. Thus $ExpTot^{\sigma'}_{\mathcal{M}}(\boldsymbol{w}\cdot\boldsymbol{\rho}[\boldsymbol{k}]) = \boldsymbol{w} \cdot \boldsymbol{p'} \geqslant \boldsymbol{w} \cdot \boldsymbol{p}$ and by the maximality of $\sigma$ we get $\boldsymbol{w} \cdot \boldsymbol{q} \geqslant \boldsymbol{w} \cdot \boldsymbol{p'}$. Then $\boldsymbol{w} \cdot \boldsymbol{q} \geqslant \boldsymbol{w} \cdot \boldsymbol{p}$.

For the second item, let $F$ be the face from Lemma 1 obtained for the vector $\boldsymbol{w}$. Suppose there is some $\boldsymbol{x} \in X \cap F$. By definition of $F$ we have $\boldsymbol{w} \cdot \boldsymbol{q} = \boldsymbol{w} \cdot \boldsymbol{x}$. Because by the definition of $\boldsymbol{w}$ and of the separating hyperplane we have $\boldsymbol{w} \cdot \boldsymbol{x} < \boldsymbol{w} \cdot \boldsymbol{p}$, we get that $\boldsymbol{w} \cdot \boldsymbol{q} < \boldsymbol{w} \cdot \boldsymbol{p}$, which is a contradiction.

Lemma 2(i) implies that when the algorithm returns "false", the answer is correct. The answer "true" is correct by the definition of separating hyperplane and the convexity of the set of achievable vectors.

Lemma 2(ii) implies that the set $X$ in the algorithm only contains points from faces. Because the number of faces is at most exponential in the size of the problem, the algorithm terminates after at most exponentially many loops.

### A.3 Correctness of Alg. 2

Let $gt(i)$ denote the set of $j \in \{1, \dots, n\}$ satisfying $k_j - 1 \geqslant i$, let $k_{max} = \max_{i \in \{1,\dots,n\}} k_i$, and let $p^{\sigma}_{s,a}(i)$ be an abbreviation for $Pr^{\sigma}_{\mathcal{M}}(\{\pi = s_0 a_0 s_1 a_1 \dots \mid s_i = s \text{ and } a_i = a\})$.

For all adversaries $\sigma \in Adv_{\mathcal{M}}$ we have:

$$ExpTot^{\sigma,\boldsymbol{k}}_{\mathcal{M}}(\boldsymbol{w}\cdot\boldsymbol{\rho}) = \sum_{i=0}^{\infty} \sum_{j \in gt(i)} w_j \cdot \sum_{s \in S, a \in \alpha} \rho_j(s,a) \cdot p^{\sigma}_{s,a}(i)$$

Let us begin by analysing the first do-while cycle. After it is performed $m$-times, we get that the value of $x_s$ is equal to:

$$\max_{\sigma \in Adv_{\mathcal{M}}} \sum_{i=0}^{m-1} \sum_{j \in gt(\infty)} w_j \cdot \sum_{s \in S, a \in \alpha} \rho_j(s,a) \cdot p^{\sigma}_{s,a}(i)$$

which can be proved by simple induction on $m$. The first do-while cycle is in fact an "ordinary" value iteration w.r.t. reward structure $\bar{\rho}$ given by $\bar{\rho}(s,a) = \sum_{j\in gt(\infty)} w_j \rho_j(s,a)$ which is either always non-negative, or always non-positive (due to the condition (iii) from page 7), and hence we can use techniques from [19] to prove that the value of $x_s$ converges to:

$$\max_{\sigma\in Adv_{\mathcal{M}}} \sum_{i=0}^{\infty} \sum_{j\in gt(\infty)} w_j \cdot \sum_{s\in S, a\in\alpha} \rho_j(s,a)\cdot p^{\sigma}_{s,a}(i)$$

Let $z_s$ be the value of $x_s$ after all the iterations of the first do-while cycle have finished. For the third do-while while cycle, the key observation for the correctness is that the value of $x_s$ after the cycle is repeated $\ell$ times is equal to:

$$\max_{\sigma\in Adv_{\mathcal{M}}} \left( \sum_{s\in S, a\in\alpha} p^{\sigma}_{s,a}(\ell)\cdot z_s \right) + \left( \sum_{i=0}^{\ell-1} \sum_{j\in gt(k_{max}-\ell+i)} \sum_{s\in S, a\in\alpha} \rho_j(s,a)\cdot p^{\sigma}_{s,a}(i) \right)$$

When $\ell = k_{max}$ and as $z_s$ goes to $ExpTot^{\sigma,\boldsymbol{k}}_{\mathcal{M}}(\boldsymbol{w}\cdot\boldsymbol{\rho})$, this number is equal to (below, $q^{\mathcal{M}(s),\sigma'}_{s',a}(i)$ stands for $Pr^{\sigma}_{\mathcal{M}(s)}(\{\pi = s_0a_0s_1a_1\ldots \mid s_i = s \text{ and } a_i = a\})$ where $\mathcal{M}(s)$ is $\mathcal{M}$ with the initial state changed to $s$):

$$\max_{\sigma\in Adv_{\mathcal{M}}} \left( \sum_{s\in S, a\in\alpha} p^{\sigma}_{s,a}(k_{max})\cdot z_s \right) + \left( \sum_{i=0}^{k_{max}-1} \sum_{j\in gt(i)} w_j \cdot \sum_{s\in S, a\in\alpha} \rho_j(s,a)\cdot p^{\sigma}_{s,a}(i) \right)$$

$$= \max_{\sigma\in Adv_{\mathcal{M}}} \left( \sum_{s\in S, a\in\alpha} p^{\sigma}_{s,a}(k_{max})\cdot \max_{\sigma'\in Adv_{\mathcal{M}}} \sum_{i=0}^{\infty} \sum_{j\in gt(\infty)} w_j \cdot \sum_{s'\in S, a'\in\alpha} \rho_j(s',a')\cdot q^{\mathcal{M}(s),\sigma'}_{s',a'}(i) \right)$$
$$+ \left( \sum_{i=0}^{k_{max}-1} \sum_{j\in gt(i)} w_j \cdot \sum_{s\in S, a\in\alpha} \rho_j(s,a)\cdot p^{\sigma}_{s,a}(i) \right)$$

$$= \max_{\sigma\in Adv_{\mathcal{M}}} \left( \max_{\sigma'\in Adv_{\mathcal{M}}} \sum_{s\in S, a\in\alpha} p^{\sigma}_{s,a}(k_{max})\cdot \sum_{i=0}^{\infty} \sum_{j\in gt(\infty)} w_j \cdot \sum_{s'\in S, a'\in\alpha} \rho_j(s',a')\cdot q^{\mathcal{M}(s),\sigma'}_{s',a'}(i) \right)$$
$$+ \left( \sum_{i=0}^{k_{max}-1} \sum_{j\in gt(i)} w_j \cdot \sum_{s\in S, a\in\alpha} \rho_j(s,a)\cdot p^{\sigma}_{s,a}(i) \right)$$

$$= \max_{\sigma\in Adv_{\mathcal{M}}} \left( \sum_{i=k_{max}}^{\infty} \sum_{j\in gt(\infty)} w_j \cdot \sum_{s\in S, a\in\alpha} \rho_j(s,a)\cdot p^{\sigma'}_{s,a}(i) \right)$$
$$+ \left( \sum_{i=0}^{k_{max}-1} \sum_{j\in gt(i)} w_j \cdot \sum_{s\in S, a\in\alpha} \rho_j(s,a)\cdot p^{\sigma}_{s,a}(i) \right)$$

$$= \sum_{i=0}^{\infty} \sum_{j\in gt(i)} w_j \cdot \sum_{s\in S, a\in\alpha} \rho_j(s,a)\cdot p^{\sigma}_{s,a}(i)$$

which gives the correctness of computation of $x_s$.

### A.4 Correctness of Alg. 3

Let us now prove the correctness of Alg. 3. The correctness of the returned value $\perp$ can be proved in the same way as proving the correctness of the value "false" returned by Alg. 1. The following lemma is a modification of Lemma 2 and gives an insight into the correctness of the algorithm for the values different to $\perp$.

**Lemma 3.** *Let* $\boldsymbol{\psi} = ([\rho_1]^{\leqslant k_1}_{\sim_1 p_1}, \ldots, [\rho_n]^{\leqslant k_n}_{\sim_n p_n})$ *be a multi-objective query, let* $X$ *be a set of vectors* $(r_1, \ldots, r_n)$ *such that* $\boldsymbol{\psi} = ([\rho_1]^{\leqslant k_1}_{\geqslant r_1}, \ldots, [\rho_n]^{\leqslant k_n}_{\geqslant r_n})$ *is achievable and* $\boldsymbol{p} = (p_1, \ldots p_n)$ *is on the boundary of* $down(X)$*, and let* $\boldsymbol{w}$ *be such that* $w_1 > 0$ *and it separates* $\boldsymbol{p}$ *from* $down(X)$*. Let* $\sigma$ *be an adversary maximising* $ExpTot^{\sigma}_{\mathcal{M}}(\boldsymbol{w}\cdot\boldsymbol{\rho}[\boldsymbol{k}])$*, and* $\boldsymbol{q} = (ExpTot^{\sigma}_{\mathcal{M}}(\rho_i))_{1\leqslant i\leqslant n}$*. Then the following is true:*

- *If* $\boldsymbol{w}\cdot\boldsymbol{q} = \boldsymbol{w}\cdot\boldsymbol{p}$*, then* $\boldsymbol{\psi}' = ([\rho_1]^{\leqslant k_1}_{\sim_1 p'}, [\rho_2]^{\leqslant k_2}_{\sim_2 p_2}, \ldots, [\rho_n]^{\leqslant k_n}_{\sim_n p_n})$ *is not satisfiable for any* $p' > p_1$*.*
- *If* $\boldsymbol{w}\cdot\boldsymbol{q} > \boldsymbol{w}\cdot\boldsymbol{p}$*, then there is a face* $F$ *(of the convex polytope of achievable solutions) such that* $\boldsymbol{q}$ *is on* $F$*, but no point from* $X$ *is in* $F$*.*

*Proof.* We start with the first item and prove it by contrapositive. If some $\boldsymbol{\psi}'$ is achievable, then there must be $\boldsymbol{p'}$ such that $p'_i \geqslant p_i$ for all $i \in \{2, \ldots, n\}$ and $p'_1 > p_1$, and there must be an adversary $\sigma'$ such that $ExpTot^{\sigma',k_i}_{\mathcal{M}}(\rho_i) = p'_i$ for all $i$. Thus $ExpTot^{\sigma'}_{\mathcal{M}}(\boldsymbol{w}\cdot\boldsymbol{\rho}[\boldsymbol{k}]) = \boldsymbol{w}\cdot\boldsymbol{p'}$ and by the maximality of $\sigma$ we get $\boldsymbol{w}\cdot\boldsymbol{q} \geqslant \boldsymbol{w}\cdot\boldsymbol{p'}$. We also have $\boldsymbol{w}\cdot\boldsymbol{p'} = \sum_{i=1}^{n} w_i \cdot p'_i > \sum_{i=1}^{n} w_i \cdot p_i = \boldsymbol{w}\cdot\boldsymbol{p}$: The inequality follows because $p'_i \geqslant p'_i$ for $i \in \{2, \ldots, n\}$ (and hence $\sum_{i=2}^{n} w_i \cdot p'_i \geqslant \sum_{i=2}^{n} w_i \cdot p_i$) and because $p'_1 > p_1$ and $w_1 > 0$, so $w'_1 \cdot p_1 > w_1 \cdot p_1$. This yields $\boldsymbol{w}\cdot\boldsymbol{q} > \boldsymbol{w}\cdot\boldsymbol{p}$.

The second item follows similarly to the corresponding part of Lemma 2.

As in the case of Alg. 1, the second item of Lemma 3 together with the fact that the number of faces is finite gives us that the algorithm terminates.

### A.5 Correctness of Alg. 4

We sketch the correctness of Alg. 4 as follows. The following invariants hold throughout its execution:

1. The set $down(X)$ contains only achievable points, and $X$ contains only points on the boundary of the set of achievable solutions.
2. No point $\boldsymbol{r}$ satisfying $\boldsymbol{w}\cdot\boldsymbol{r} > \boldsymbol{w}\cdot\boldsymbol{q}$ for some $\boldsymbol{q} \in X$ and $\boldsymbol{w} \in Y(\boldsymbol{q})$ is achievable.

Let us assume the algorithm has chosen the point $\boldsymbol{p}$ for points $\boldsymbol{x}^i$, $\boldsymbol{x}^{i+1}$. The vector $\boldsymbol{w}$ is chosen so that one of the following happens:

- A point $\boldsymbol{q}$ is found such that $\boldsymbol{w}\cdot\boldsymbol{q} > \boldsymbol{w}\cdot\boldsymbol{x}^i = \boldsymbol{w}\cdot\boldsymbol{x}^{i+1}$.
- A point $\boldsymbol{q}$ is found such that $\boldsymbol{w}\cdot\boldsymbol{q} = \boldsymbol{w}\cdot\boldsymbol{x}^i = \boldsymbol{w}\cdot\boldsymbol{x}^{i+1}$

In the first case, a new face is found. In the second case, $\boldsymbol{w}$ is added to $Y(\boldsymbol{q})$ and due to the way the points $\boldsymbol{p}$ are chosen, no point $\boldsymbol{p}$ satisfying $x^i_1 \leqslant p_1 \leqslant x^{i+1}_1$ will be chosen in the future. This implies that the algorithm terminates.

### A.6 Further Details for Implementation Optimisations

Our value iteration-based method (Alg. 2) employs the Gauss-Seidel variant of value iteration [19]: instead of using a pair of vectors $\boldsymbol{x}$ and $\boldsymbol{y}$ to store the solution vector a single vector $\boldsymbol{x}$ is used. Its elements are then updated as soon as they are computed. This increases the speed of convergence and also decreases memory requirements. Furthermore, storage of vectors can then be optimised to save space as follows: We can choose not to store the $2n$ vectors $\boldsymbol{x}_i$ and $\boldsymbol{y}_i$ in Alg. 2, but instead use a single pair $\boldsymbol{x}'$ and $\boldsymbol{y}'$ and compute the $n$ objectives separately. Alternatively, we can always omit storage (and computation) of one of the individual objectives, noting that it can be reconstructed from the weighted objective and the other $n-1$ individual ones. This can be significant when $n$ is small, which is often the case in practice.